# A Workflow-Centric Approach to Generating FAIR Data Objects for Computationally Generated Microstructure-Sensitive Mechanical Data

*Ronak Shoghi* *, *Alexander Hartmaier*

R. Shoghi, A. Hartmaier

ICAMS, Ruhr-Universität Bochum, Universitätsstraße 150, 44801, Bochum, Germany

E-mail: ronak.shoghi@rub.de

E-mail: alexander.hartmaier@rub.de

## Abstract

From a data perspective, the materials mechanics field is characterized by sparsity of available data, mainly due to the strong microstructure-sensitivity of properties like strength, fracture toughness, and fatigue limit. This requires testing specimens with different thermo-mechanical histories, even when the composition is similar. Experimental data on mechanical behavior is rare, as mechanical testing is destructive and requires significant material and effort. Furthermore, mechanical behavior is typically characterized in simplified tests under uniaxial loading conditions, whereas a complete characterization requires multiaxial testing. To address this data sparsity, simulation methods like micromechanical modeling can contribute to microstructure-sensitive data collections. This work introduces a novel data schema integrating both metadata and mechanical data, following the workflows of the material modeling processes by which the data has been generated. Each workflow run produces unique data objects by incorporating user, system, and job-specific information correlated with mechanical properties. This approach can be applied to any type of workflow as long as it is well-defined. This integrated format provides a sustainable way of generating Findable, Accessible, Interoperable, and Reusable (FAIR) data objects. The metadata elements focus on key features required to characterize microstructure-specific data, simplifying the collection of purpose-specific datasets by search algorithms.

## 1. Introduction

Recent progress in data-centric approaches in various scientific disciplines has profoundly transformed the process of scientific discovery and the implementation of technological innovations. In this context, sustainability has become an important concept, and effective data management is a practice of achieving sustainable scientific research. This approach enables systematic storage, organization, and access to research data. On an individual level, effective data management allows for the validation of findings, reproduction of results, and further developments based on existing works. On a larger scale and from a broader perspective, such practices ensure that data can be shared and reused collaboratively within the scientific community. Moreover, effective data management is a requirement for open science. Ruben et al. [1] described open science as a model of shared, collaborative knowledge that is transparent and accessible. In this context, adhering to FAIR principles — Findability, Accessibility, Interoperability, and Reusability — is of great importance [2]. However, despite the apparent importance of effective data management, many scientific domains still lack the tailored tools and workflows that specifically address the requirement of transforming project or user-specific data and its characteristics into FAIR data objects. Without proper documentation, extending the data lifecycle and ensuring sustainable research would not be possible. A key element of these required tools and workflows is the metadata schema or – more generally – a data standard. Metadata or *data about data* is characterized by Smiraglia [3] as structured and encoded elements that describe characteristics of an information-bearing entity. These elements aid in identifying, discovering, assessing, and managing the described entity [4]. A metadata schema or data standard specifies the structure and elements of metadata, ensuring a uniform description and structure [5]. Well-defined metadata schema promotes data interpretability and reusability, supporting an extended data life cycle. Such a metadata schema is essential in effective data management and supports sustainable research practices at both individual and global levels.

While general metadata schemas such as Dublin Core [6], DataCite [7], and the Materials Object Description Schema (MODS) [8] are used for a variety of data types and provide foundational metadata elements, they often fall short when addressing unique requirements of domain-specific data types. These general schemas can capture basic and general aspects of data, but the complexity of the domain-specific data necessitates more tailored metadata schemas that can handle the detailed and specialized layers of information. In the field of material science, domain-specific schemas have been developed in recent years. Schmitz et al. [9] suggested a schema for the geometric description of 3D microstructures. This schema suggests a comprehensive system of descriptors to describe static microstructures' spatial and compositional characteristics independent of any specific numerical model. Later, Reith et al. [10] expanded the schema with elements to describe the atomistic simulation data in a job-centric structure. Joseph et al. [11] proposed a metadata schema tailored for scanning electron microscopy (SEM) measurements. This schema addresses the challenges posed by various metadata standards from different manufacturers and the complexities of correlating data from multiple instruments. The suggested schema facilitates interpretability and data exchange. Li et al. [12] suggested a metadata architecture for additive manufacturing (AM) projects to capture five levels of information: project management, feedstock materials, AM building and post-processing, microstructure and properties measurements, and computer simulations. The suggested schema can facilitate the integration of datasets from different sources. Rao et al. [13] suggested a metadata schema for lattice thermal conductivity from first principle calculations which is divided into three consecutive processes. For each process, a detailed metadata schema has been proposed with grouped metadata element sets.

Up to this point, most metadata schemas in the materials science domain have focused on atomic level or microstructural data, while data on mechanical material behavior has not been widely addressed. The reason for this may lie in the severe microstructure-sensitivity and history dependence of mechanical materials data, where different thermo-mechanical process routes lead to significant differences in mechanical properties such as strength, fracture toughness, and fatigue limit. To address this history dependency, we suggest incorporating the entire workflow of materials processing up to the point of data collection in the data standard. Such a workflow-centric approach requires an integrated data standard in which metadata and mechanical data form one single data object. Furthermore, sustainability in this context requires a strict limitation of user interference to reduce the burden on users to provide information on the process history that needs to be collected. Such limitations in user efforts will finally increase the acceptance of a data standard. As a first step toward establishing such an integrated approach, the focus is placed on creating a data standard for mechanical material data from numerical simulations. Numerical simulations allow for the generation of large data sets for different mechanical loading scenarios, whereas experiments are typically restricted to uniaxial testing of a small number of specimens due to the great effort in specimen preparation and testing. Furthermore, in numerical simulations, the workflows from microstructure generation to simulation of the mechanical properties are well-controlled and typically lie in the hands of a single user, which facilitates the automized data collection of the relevant microstructure data. Hence, we propose an integrated data standard designed explicitly for micromechanical simulations, which can be generalized in subsequent steps to more complex workflows, including simulations of thermo-mechanical processing of the material prior to mechanical testing and finally even to experimental workflows, which will pose the highest requirements on such an integrated data standard. In this workflow-centric approach, we treat each workflow instance — including user and system information, material parameters, microstructure descriptors, and resulting mechanical behavior — as a unique data object. Our objective is to design a metadata schema that captures various levels of available information essential for the proper description of this data object, ensuring its adherence to FAIR principles. The schema is structured to detail every aspect of the workflow, from user interactions to system configurations, job-specific parameters, and properties correlated to the specific material model definition and the resulting mechanical properties. This approach can be applied to any well-defined workflow. The example presented here demonstrates the applicability of this method for computational workflows, but the underlying concept serves as a general guideline for structuring metadata in any process. Whether in a simulation or an experiment, following a well-defined workflow — including user interactions, system configurations (either a computational environment or an experimental setup), and process-specific steps— provides the framework to capture all necessary metadata elements. Ultimately, this offers a straightforward way to define FAIR data objects and contributes to sustainable research data management.

## 2. Definition of data object

Micromechanical modeling integrates the microstructural characteristics of materials into predictions of macroscopic behavior. This multiscale approach does not treat the material as a homogeneous entity but aims to explicitly incorporate the effect of microconstituent and their interaction. On the microscale, the infinitesimal material neighborhood is not uniform and consists of various constituents with different properties. The Representative Volume Element (RVE) plays an important role within this framework. The RVE, a small statistically representative sample of the material microstructure, is used to derive continuum constitutive properties by averaging the behavior over the microscale and predicting the overall behavior of the material at the macro-scale using various homogenization techniques [14]–[16].

In practice, different computational methods and tools can be used for micromechanical simulations. However, independent of the software and method employed, these simulations typically consist of the following steps:

1. *Microstructure definition:* characterization of material's internal features such as RVE definition, grain orientation descriptions, sizes, and shapes.
2. *Material models:* specification of constitutive models that describe the material behavior, including elasticity, plasticity, and damage mechanisms.
3. *Boundary conditions:* defining the relevant boundary conditions and loading scenarios that reflect the real-world working conditions of the material
4. *Solver settings:* parameters and numerical methods used to perform the simulation and solving of the governing equations

These specific *tasks* or *steps* lead to a description of *a job* in computational terms. A job can be defined as a complete unit of work submitted for execution. However, handling a job effectively requires a workflow. A workflow involves managing, coordinating, and executing one or multiple jobs in a structured manner and usually includes a post-processing step to derive the desired mechanical properties. A workflow can be utilized by any user on any system. Ideally, a workflow should be connected to a database. As shown in **Figure 1**, a workflow instance represents a single occurrence of the workflow. Each instance is unique as it has been executed by user $U$, on system $S$ with the selection of the parameters that define the job $J$, resulting in mechanical properties $P$. This unique instance is the *data object* and is then transferred to a database. We aim to define the representative elements — metadata — of this data object.

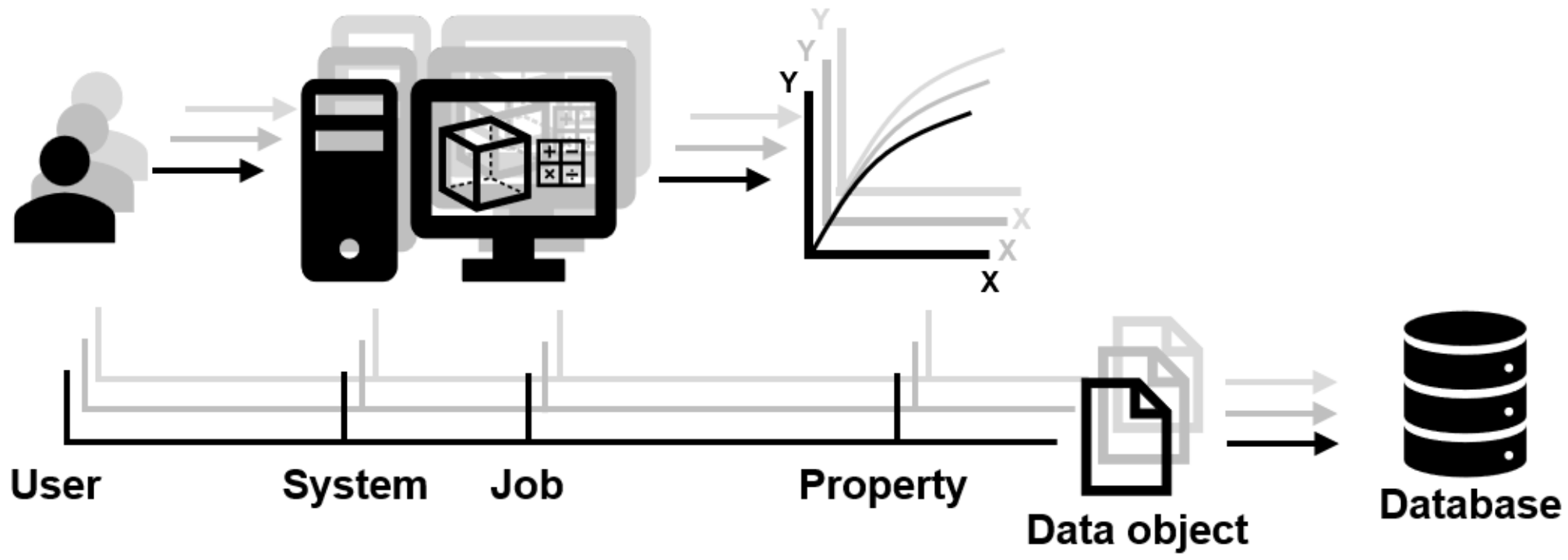

**Figure 1.** Illustration of the end-to-end process of a micromechanical modeling workflow. It highlights the flow from user initiation to database storage, emphasizing the uniqueness and structure of each workflow instance. Key components include a **User** who initiates the workflow. The user can select both the system and the job parameters, enabling a customized execution path. The **System** symbolizes the computation environment chosen by the user. It includes specific hardware specifications and software environments, varying depending on the user's preferences and job requirements. A **Job** is defined by the user. This definition involves setting the parameters that dictate the execution of the computational task, tailored to derive desired results. The **Property** represents the mechanical properties, which are outcomes of the job. Depending on the workflow's structure, these properties may have already undergone a post-processing step. The **Data object** includes all relevant information from the user, system, job parameters, and the resulting properties, which is transferred to a database. The **Database** is the destination for the data objects, ensuring systematic storage and future accessibility for analysis and reference. Additionally, this figure includes multiple instances of the workflow, shown in varying shades. Each shade represents a unique instance of the workflow.

Figure 1 also includes instances shown in different shades to illustrate multiple instances of the workflow conceptually. These shaded instances represent other workflow executions, each distinctively configured by different users on different systems or with varying selections of job-related parameters. This visualization emphasizes the concept of multiple instances to demonstrate the systematic and consistent approach of the workflow, regardless of variations in execution parameters. This workflow-centric approach supports efficient data management by enabling systematic organization of data. It ensures that data or specific properties are findable and accessible through effective search and filtration based on metadata elements. Moreover, this approach facilitates reproducibility by maintaining a complete record of simulation parameters and conditions, supports scalability by easily integrating additional jobs, and enhances collaboration by providing a clear structure for data sharing.

## 3. Metadata schema design for the data object

The described data object contains various levels of informational components, each differing in nature. These components may have different roles in the existence of data objects and can also have different functionalities. All the potential metadata elements can be classified into various categories and hierarchies. Angermann et al. [17] defined taxonomy as classifying objects with similar properties into subcategories. However, as explained by Lambe [18], this classification is not solely based on similarity attributes but also based on different kinds of relationships. In this schema, the classification of the data object is based on the level of specificity. It is important to capture each level of information to ensure that all aspects of the data object—from the most specific level to the highest general level—are organized and aligned with FAIR principles. As shown in **Figure 2**, there are four levels of available information on which the elements can be described. These categories are summarized as follows:

- **User-specific elements**: This level of information contains user-specific details that make the data object unique. The general elements align with those typically found in standard metadata schemas such as Dublin Core [6] or DataCite [7]. Essential elements at this level include information about the creator or owner, their affiliation, date, and details about ownership and licenses. These elements are crucial for sharing and collaboration.
- **System elements**: This level of information focuses on capturing technical and system-related information critical for replicating the simulation environment. It includes information about computational systems, software versions, and file paths. Including this information is of great importance as it can ensure the reproducibility of the simulation setup and the data object.
- **Job-specific elements**: The most detailed level which deals directly with the simulation setup. This level of information is usually defined by the user in the input file of the simulation and can be divided into three subcategories: *Geometry*, *Material model,* and *Boundary condition.*
- **Property elements:** This category includes elements describing the outputs of each simulation unit. Usually, the output data undergoes a post-processing step to transform it into the desired format. In the case of micromechanical simulations, these outputs are in the form of mechanical responses or properties with respect to the specifications defined through the model-level elements.

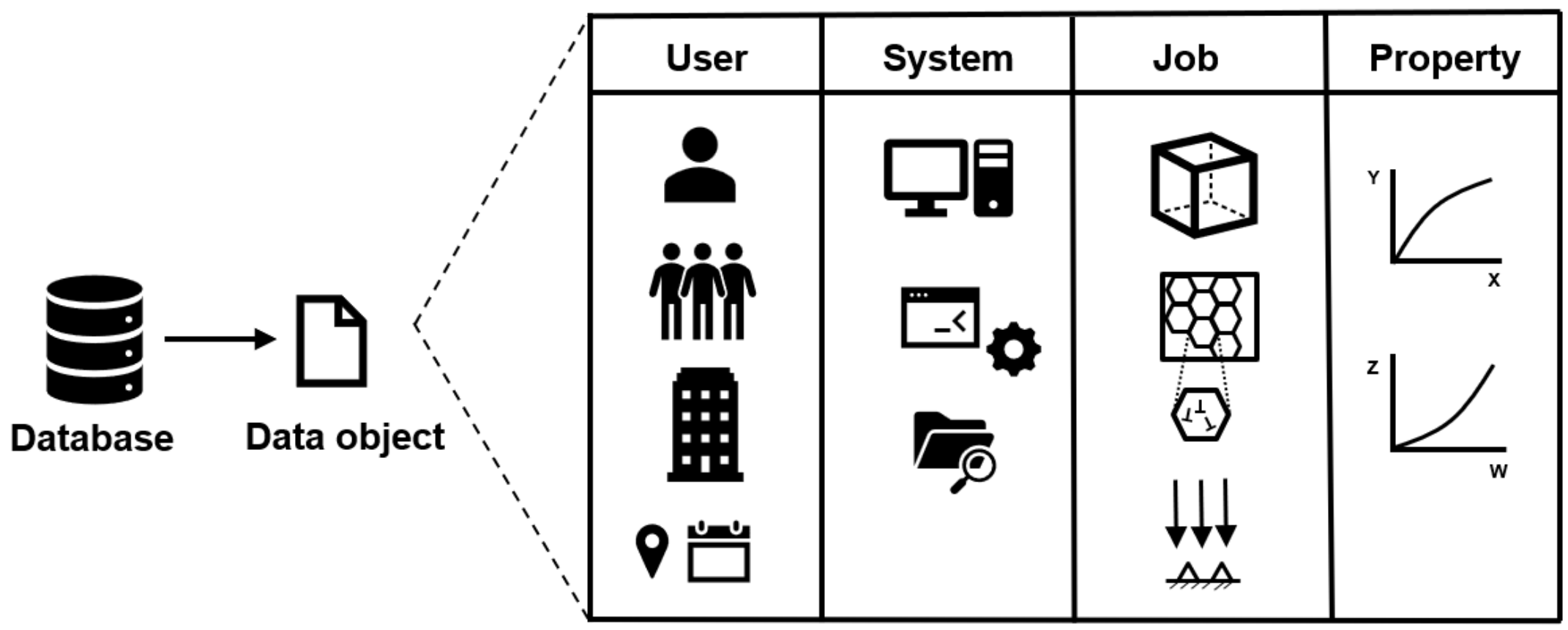

**Figure 2**. Different levels of available information can be used to design the elements of metadata schema. This information can be categorized into four levels. **User**-specific information includes essential user-specific details such as identity, affiliations, ownership, and date. **System**-level information includes computational environment specifics such as hardware and software configurations. **Job**-related information, which is the most specific level, is provided by the user. This includes specific parameters that describe the job, such as elements defining the geometry, material model, and boundary conditions. **Property**-level information includes different mechanical properties derived from the job's output, typically post-processed within the workflow.

As mentioned, the metadata schema or standard provides a structured framework that defines how data can be described through various elements [5], [19]. With the categories defined previously, we aim to define the schema and organize the endless ocean of information into comprehensive categories through metadata elements. Designing the metadata elements or choosing *the good enough elements* is a critical task for effectively capturing the data object's characteristics. As described by Zeng et al. [19], it is vital to avoid over-designing the metadata schema with too many elements, which can violate the universal principle of *simplicity*. Excessive elements can complicate the schema unnecessarily, making it challenging to implement and maintain. While simplicity is fundamental, it is also essential to have a flexible schema that can be extended to include new elements or sub-elements. This *extensibility* allows the schema to remain relevant and effective, promoting sustainability, an important principle that should always be considered. An element set should be designed for long-term *applicability*, not just short-term adoption within a specific project, and should be created with the future in mind [19]. After designing the elements, defining descriptive vocabularies is crucial to ensure clarity and interoperability for data management. Using existing element names whenever possible is important, which is usually the case for general elements. Introducing new names requires explicit definitions and alignments (crosswalks) with widely accepted elements in schemas like Dublin Core [6]. While creating new elements in domain-specific schemas is necessary, generic schemas cannot capture the rich discipline-specific descriptions of data. In this case, employing accepted domain-specific terminology is crucial. It provides the basis for standardized definitions of elements and ensures conceptual alignment across the domain [20]. There are general naming conventions, as highlighted in [21], which emphasize that names should be specific and descriptive. Additionally, as explained in [20], names must be understandable and actionable by both humans and machines. In specific domains such as materials science, additional conventions that reflect common practices within the field apply. For instance, *RVE* is a widely recognized abbreviation and can be appropriately used according to the naming conventions of this domain. This ensures the community's interoperability of data elements. When necessary, as described by Zheng et al. [22], a new vocabulary for an element can be derived based on methods such as adaptation, expansion, and modification using existing terms as a *source* or *model* to ensure semantic interpretability.

## 4. Metadata schema: detailed structure and elements

This section outlines the detailed elements categorized within the metadata schema, distinguishing between Mandatory (M) and Optional (O) elements. The formatting convention employed here is *snake case*, aligning with Python coding standards. Where feasible, particularly for user-specific and system-related elements, crosswalks and mappings to generally accepted schemas have been done to ensure compatibility and integration.

### 4.1. User-specific elements

These elements originate from the most general level of available information and are fundamentally connected to the user in two distinct ways:

- **Pre-existing (known) information**: some elements in this category are intrinsically linked to the inherent identity and context of the user or creator. For instance, elements such as creator name, creator affiliation, and the date the data object has been created fall into this category. These elements do not usually require an active decision by the user; they are often automatically extracted based on the system's understanding of the user's identity and context.
- **Decision-required information:** in contrast, some elements require explicit decisions and actions from the user. These decision-required elements include titles, descriptions, and rights management. Additionally, decisions about identifying and attributing contributors are critical and require user intervention and decision.

These user-specific elements are summarized in **Table 1**. It is important to emphasize that including these elements for each data object is essential to ensure adherence to FAIR principles. In practice, there are scenarios where individual data objects are retrieved and compiled into collections through search and filtration. In this case, each data object within the collection must hold all necessary information to ensure the collection is informative enough for usage or sharing purposes. Furthermore, the assembly of such a collection is only possible when the sharing permissions within the data objects match, indicating that they belong to the same authorized group. This is typically managed through specific metadata elements such as *shared_with*. This explicit user consent not only adheres to legal and ethical standards but also enhances the interoperability and reusability of the data. It is important to note that since this level of information is inherently connected to the user, the automatic collection of these elements is generally straightforward. Pre-existing information can typically be defined by the user during their initial use of the workflow, while elements requiring a user decision can be collected through user intervention within the automated workflow. A critical element in defining a FAIR data object, and maybe the most crucial element in this information level, is the unique identifier or name of the data object. The existence of this identifier is vital to ensure that the data can be referenced and accessed effectively. There are different methods for generating this unique identifier within an automated workflow, which will be explained in Section 5.

**Table 1.** User-specific elements, obligation choice, type, and crosswalk to DataCite and Dublin Core schemas.

| Elements | Obl | Type | DataCite | Dublin Core | Description |
|---|---|---|---|---|---|
| identifier | M | Str | identifier | identifier | Name of the data object, serving as the unique identifier. |
| title | M | Str | titles.title | title | The title of the data object or the project where the data object is part of it. |
| creator | M | List of Str | creators.creator.creatorName | creator | The user responsible for generating the data object, formatted as "Last name, First name". Multiple creators are inserted as a list. |
| creator_ORCID | O | List of Str | creators.creator.nameIdentifier | creator | The ORCID ID of the creator. Multiple ORCID IDs are inserted as a list. |
| creator_affiliation | M | List of Str | creators.creator.affiliation | creator | The organizational affiliation of the creator. Multiple affiliations are inserted as a list. |
| creator_institute | O | List of Str | - | creator | The specific institute within the organization to which the creator belongs. Multiple institutes are inserted as a list. |
| creator_group | O | List of Str | - | creator | The group within the organization associated with the creator. Multiple groups are inserted as a list. |
| contributor | O | List of Str | contributors.contributor.contributorName | contributor | Individuals who have contributed significantly to the data generation process, formatted as "Last name, First name". Multiple creators are inserted as a list. |
| contributor_ORCID | O | List of Str | contributors. contributor.nameIdentifier | contributor | The ORCID ID of the contributor. Multiple ORCID IDs are separated by commas. |
| contributor_affiliation | O | List of Str | contributors. contributor.affiliation | contributor | The organizational affiliation of the contributor. Multiple affiliations are inserted as a list. |
| contributor_institute | O | List of Str | - | contributor | The specific institute within the organization to which the contributor belongs. Multiple institutes are inserted as a list. |
| contributor_group | O | List of Str | - | contributor | The group within the organization associated with the contributor. Multiple groups are inserted as a list. |
| date | M | Str | dates.date | date | The date when the data generation process started in the format YYYY-MM-DD. |
| shared_with | M | List of Obj | - | - | Specifies who can access the data object. The categories can be c (Creator): The creator of the data. u(User): Specific users g (Group): The specific group of individuals. all: Public access. See **Appendix 2.** |
| description | O | Str | descriptions.description | description | A free-text description of the data object. |
| rights | M | Str | rightsList.rights | rights | The license under which the dataset is made available. |
| rights_holder | M | List of Str | - | rightsHolder | The person or organization managing rights over the resource. |
| funder_name | O | Str | fundingReferences. fundingReference.funderName | - | The name of the funding provider. |
| fund_identifier | O | Str | fundingReferences. fundingReference.awardNumber | - | The project identifier provided by the funding agency or body. |
| publisher | O | Str | publisher | publisher | The entity responsible for making the dataset available. |
| relation | O | List of Str | RelatedIdentifier | relation | Any resource related to or associated with the dataset. |
| user_extra_information | O | Str | - | - | Additional user details or information relevant to the data object. |
| keywords | O | List of Str | Subjects.subject | subject | Words that highlight the main topics or concepts of the data object separated by commas. |

### 4.2. System elements

Numerical simulations are inherently deterministic in nature. Given a set of initial parameters (such as those required to define the job), the outcome of the simulation is uniquely determined by the governing equations and the specific numerical methods used to solve them. However, practical experience and recent studies have highlighted reproducibility as a significant challenge in simulation-based experiments, even under controlled conditions [23], [24]. Different solvers, such as Finite Element Method (FEM) or Fast Fourier Transformation (FFT), use different numerical algorithms to address these governing equations. This naturally leads to variations in the results when switching between different solvers. However, different results can still be obtained even when using the same solver and an identical set of initial parameters. Such variations can arise from factors like floating point precision, compiler differences, hardware architecture, or parallel computing [23], [24]. To address these challenges, including system-related information as part of the metadata schema is crucial. This includes not only the software and its version but also the operating system and its version and the processor's specifications. Additionally, the metadata should include the physical storage locations (file paths) of the input and output files associated with each data object or workflow instance. By including these elements, it becomes feasible to reproduce the data object without redundant computation. While model-related elements should enable reproduction without relying on the input file, having access to the input file can save time and resources. To protect sensitive information, it is important to use relative paths that describe the location of files in relation to the project's root directory or anonymized details such as usernames or server names with generic placeholders or abstract identifiers. For the technical or system-level metadata, there are no specified terms in general schemas like DataCite [7] or Dublin Core [6]. In this case, the CodeMeta [25] schema can be used for mapping and crosswalk. The detailed system-specific elements can be seen in **Table 2**.

**Table 2.** System-specific elements, obligation choice, type and crosswalk to CodeMeta.

| Elements | Obl | Type | CodeMeta | Description |
|---|---|---|---|---|
| software | M | Str | - | Names of the software or computational tool used for running the simulation. |
| software_version | M | Str | softwareVersion | Specifies the version number of the software or computational tool used for running the simulation. |
| system | M | Str | operatingSystem | The name of the operating system where the data was generated. |
| system_version | M | Str | - | Specifies the version of the operating system. |
| processor_specifications | M | Str | - | The technical characteristics of the processor. |
| input_path | M | Str | inputFile | The relative file path or directory where input data files required to run the simulation or analysis are stored. |
| results_path | M | Str | outputFile | The file path or directory where the results of the simulation or analysis are stored. |
| system_extra_information | O | Str | description | Additional system-related details or information relevant to the data object. |

### 4.3. Job-specific elements

The job-specific elements provide the most specific level of information required to run a simulation, typically defined by a user. We organize information based on a functional taxonomy in the job-level metadata section. A taxonomy in this context can be understood as a hierarchy of controlled vocabulary values that are used to classify content [26]. This taxonomy aims to design elements into three essential functional categories necessary to describe the simulation setup: *Geometry*, *Material model*, *and Boundary condition*.

### 4.3.1. Geometry

As mentioned earlier, in the micromechanical simulations, the aim is to predict the overall mechanical response of the material by looking at the microscale and defining the RVE, which statistically represents the microstructural features. There are different numerical strategies to solve the boundary value problem within this RVE, usually by using discretization techniques. In FEM, the body is discretized into finite elements, providing greater flexibility for handling complex geometries. Alternatively, FFT-based solvers use discretization methods on regular voxel grids [27], [28]. Independent of the method, this discretization determines the resolution of the simulation, characterized by the size and number of discrete units within the RVE. It is important to include elements describing this geometric information in the metadata to have a FAIR data object. Including these elements enables the combination of data objects to form a coherent data collection and facilitates collaborative data sharing, even when using different workflows and software. The elements in this subcategory are summarized in **Table 3**. It must be noted that there is no crosswalk for these specific metadata elements. To ensure clarity and consistency in these metadata elements, we include a *property* column that helps to specify the units used for different attributes.

**Table 3.** Geometry elements, obligation choice, type, and property

<table>
<tr><th>Components</th><th>Obl</th><th>Type</th><th>Property</th><th>Description</th></tr>
<tr><td>RVE_size</td><td>M</td><td>Array of Float</td><td>Length</td><td>Represents the dimensions of the RVE in the form: length, width, height</td></tr>
<tr><td>RVE_continuity</td><td>M</td><td>Bool</td><td>-</td><td>Describes whether the RVE has periodic or non-periodic boundary conditions, controlled values: "True" for periodic and or "False" for non-periodic</td></tr>
<tr><td>discretization_type</td><td>M</td><td>Str</td><td>-</td><td>"Structured": Indicates that the discretization uses a regular, grid-like pattern. Usually used in FFT (with voxel grid) and FEM with structured mesh.<br>"Unstructured": Indicates that the discretization uses an irregular pattern, applicable for FEM when using an unstructured mesh.</td></tr>
<tr><td>discretization_unit_size</td><td>M</td><td>Array of Float</td><td>Length</td><td>Specifies the average physical size of discrete units within the RVE, determining the resolution and level of detail.</td></tr>
<tr><td>discretization_count</td><td>M</td><td>Int</td><td>-</td><td>Total number of discrete units in the RVE</td></tr>
<tr><td>solid_volume_fraction</td><td>O</td><td>Int</td><td>-</td><td>The solid volume fraction of the microstructure, as a percentage.</td></tr>
<tr><td>origin</td><td>O</td><td>Obj</td><td>-</td><td>Information about the tool used to generate geometry.
<table>
<tr><th>Components</th><th>Obl</th><th>Type</th></tr>
<tr><td>software</td><td>O</td><td>Str</td></tr>
<tr><td>software_version</td><td>O</td><td>Str</td></tr>
<tr><td>system</td><td>O</td><td>Str</td></tr>
<tr><td>system_version</td><td>O</td><td>Str</td></tr>
<tr><td>input_path</td><td>O</td><td>Str</td></tr>
<tr><td>results_path</td><td>O</td><td>Str</td></tr>
</table></td></tr>
</table>

### 4.3.2. Boundary conditions

In micromechanical modeling, boundary conditions define how the model behaves under external influences and interacts with its surrounding environment. These conditions are crucial for setting up and solving boundary value problems, which can result in a unique solution depending on the defined boundary conditions. Boundary conditions can be specified in terms of applied external forces or stresses, which is useful when precise control of loading conditions is required. Alternatively, displacements can be prescribed at specific boundaries, ideal for scenarios where controlling the deformation path is necessary to achieve a specific strain or displacement in the material [29]. When considering an RVE, Dirichlet or Neumann boundary conditions might result in non-realistic deformation behavior [30], and usually, the periodic boundary condition (PBC) is a better choice [31]. Also, many FFT-based solvers are naturally restricted to periodic boundaries because the Fourier transform inherently assumes the data it processes is periodic. This assumption simplifies the mathematical framework and increases the computational efficiency [32], [33]. With the choice of PBC, it is possible to apply either displacement or force boundary conditions, as explained earlier. For a given RVE, displacement boundary conditions can be interpreted as an imposed macroscopic strain state relative to the RVE's edge length. On the other hand, force boundary conditions can be converted into an imposed macroscopic stress state by taking into account the RVE's cross-sectional areas [31].

In the simplest and most general case, the RVE has an orthorhombic shape and is defined by its eight vertices and six faces. Given that the size of the RVE is a known value, it is crucial to establish a systematic and straightforward method for capturing the boundary conditions applied to the RVE. Given the RVE size $lx$, $ly$, and $lz$, each vertex is named using its coordinate in the format $V_{xyz}$ as shown in **Figure 3**. The detailed description can be seen in **Table 4**.

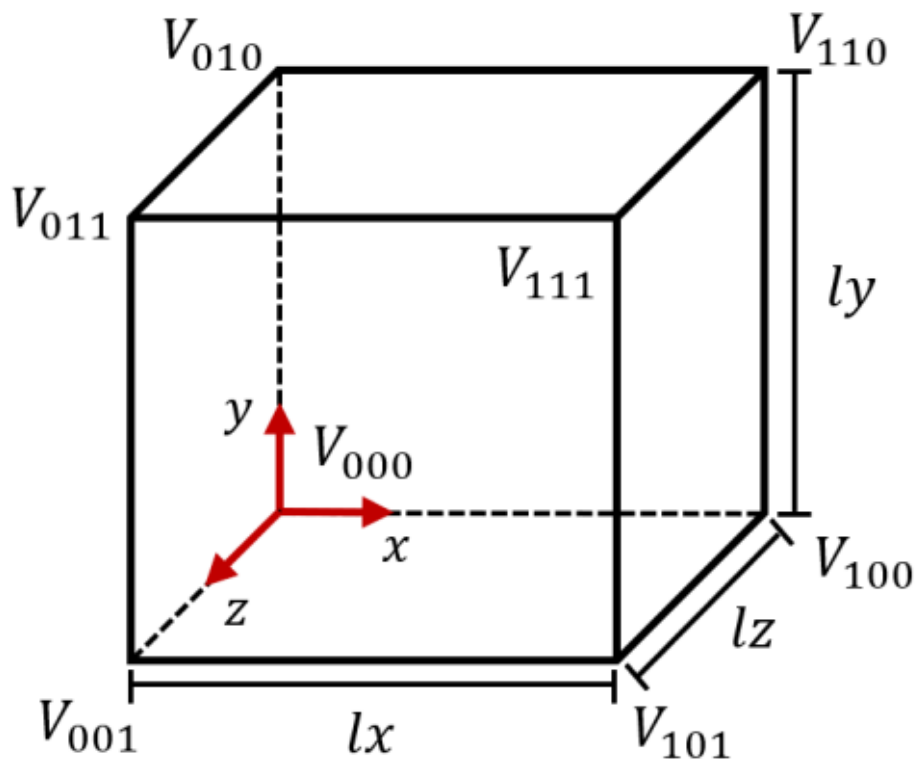

**Figure 3**. Representation of the RVE with named vertices. The RVE is shown with its eight representative vertices labeled using a systematic naming convention in which vertices are labeled based on their coordinate in $V_{xyz}$ format. The size of the RVE, captured by the dimensions $lx$, $ly$ and $lz$ is reflected in the coordinates of the vertices, and clearly indicates their positions.

**Table 4.** RVE vertices naming convention based on their coordinates, providing a clear indication of each vertex's position within the RVE

| Vertex Name | Coordinate | Description |
|---|---|---|
| $V_{000}$ | $(0,0,0)$ | Origin |
| $V_{100}$ | $(lx, 0,0)$ | Along the positive x-axis |
| $V_{110}$ | $(lx, ly, 0)$ | Along the positive x and y-axes |
| $V_{010}$ | $(0, ly, 0)$ | Along the positive y-axis |
| $V_{001}$ | $(0,0, ly)$ | Along the positive z-axis |
| $V_{101}$ | $(lx, 0, lz)$ | Along the positive x and z-axes |
| $V_{111}$ | $(lx, ly, lz)$ | Along the positive x, y, and z-axes |
| $V_{011}$ | $(0, ly, lz)$ | Along the positive y and z-axes |

Using the representative vertices set, any edge or face of the RVE can be described. Each vertex can be part of multiple sets, where a set of two vertices describes an edge and a set of four vertices describes a face. For each set, whether it contains a single vertex or multiple vertices (describing an edge or a face), the boundary condition must be described in three directions (x, y, and z).

Mechanical and thermal boundary conditions can be mixed or applied independently within micromechanical modeling. The temperature distribution or heat flux may be specified for thermal boundary conditions. For mechanical boundary conditions, constraints indicate whether a direction is free, fixed, or loaded. If a direction is free, movement in that direction is possible. If it is fixed, no movement is allowed. If it is prescribed, a specific displacement or force is applied. The boundary condition can be free, fixed, or have a specified value (force, displacement, stress or strain) for each direction. **Figure 4** shows two different examples of mechanical boundary conditions applied to an RVE. Figure 4(a) shows periodic mechanical boundary conditions applied to representative vertices of an RVE under simple tensile loading in the y-direction, where the vertex list includes only a single vertex. Figure 4(b) demonstrates mechanical boundary conditions applied to an RVE under simple shear loading in the x-direction, with two faces normal to the y-direction fixed and subjected to loading, and one edge parallel to the z-direction fixed. In this case, a list containing four vertices represents a face, while a list containing two vertices represents an edge. The detailed elements describing the boundary conditions can be seen in **Table 5**.

**Table 5.** Boundary condition elements, obligation, property, and description

<table>
<tr><th>Components</th><th>Obl</th><th>Type</th><th>Property</th><th>Description</th></tr>
<tr><td>global_temperature</td><td>O</td><td>Int</td><td>Temperature</td><td>The global initial temperature.</td></tr>
<tr><td>mechanical_BC</td><td>M</td><td>Array of Obj</td><td>-</td><td>
<table>
<tr><th>Components</th><th>Obl</th><th>Type</th><th>Property</th><th>Description</th></tr>
<tr><td>vertex_list</td><td>M</td><td>Array of Str</td><td>-</td><td>A list includes vertex/vertices that describe a point, edge, or face.</td></tr>
<tr><td>constraints</td><td>M</td><td>Array of Str</td><td>-</td><td>Defines the constraints for each direction (x, y, z) as either "free", "fixed", or "loaded".</td></tr>
<tr><td>loading_type</td><td>O</td><td>Str</td><td>-</td><td>Define the type of load applied. It can be "force", "displacement", "stress", "strain" and "none".</td></tr>
<tr><td>loading_mode</td><td>O</td><td>Str</td><td>-</td><td>Specify the nature of the loading condition with respect to time. It can be "static" (constant over time), "cyclic" (repeating periodically), "monotonic" (increasing without reversal), "intermittent" (irregular variation with time).</td></tr>
<tr><td>applied_load</td><td>O</td><td>Array of Obj</td><td>-</td><td>Contains the values of the applied load in the specified directions in constraint.
<table>
<tr><th>Components</th><th>Obl</th><th>Type</th><th>Property</th></tr>
<tr><td>magnitude</td><td>M</td><td>Int</td><td>Force/Displacement/Stress/Strain</td></tr>
<tr><td>frequency</td><td>O</td><td>Int</td><td>Frequency</td></tr>
<tr><td>duration</td><td>O</td><td>Int</td><td>Time</td></tr>
<tr><td>R</td><td>O</td><td>Int</td><td>-</td></tr>
</table>
</td></tr>
</table>
</td></tr>
<tr><td>thermal_BC</td><td>O</td><td>Array of Obj</td><td>-</td><td>
<table>
<tr><th>Components</th><th>Obl</th><th>Type</th><th>Property</th><th>Description</th></tr>
<tr><td>vertex_list</td><td>M</td><td>Array of Str</td><td>-</td><td>A list includes vertex/vertices that describe a point, edge, or face.</td></tr>
<tr><td>constraints</td><td>M</td><td>Array of Str</td><td>-</td><td>Defines the constraints for each direction (x, y, z) as either "free", "fixed", "loaded".</td></tr>
<tr><td>loading_mode</td><td>O</td><td>Str</td><td>-</td><td>Specify the nature of the loading condition with respect to time. Can be "static" (constant over time), "cyclic" (repeating periodically), "monotonic" (increasing without reversal), "fluctuated" (irregular variation with time).</td></tr>
<tr><td>applied_load</td><td>O</td><td>Array of Str</td><td>-</td><td>Contains the values of the applied load in the specified directions in constraint.
<table>
<tr><th>Components</th><th>Obl</th><th>Type</th><th>Property</th></tr>
<tr><td>magnitude</td><td>M</td><td>Int</td><td>Temperature</td></tr>
<tr><td>frequency</td><td>O</td><td>Int</td><td>Frequency</td></tr>
<tr><td>duration</td><td>O</td><td>Int</td><td>Time</td></tr>
</table>
</td></tr>
</table>
</td></tr>
</table>

(a)

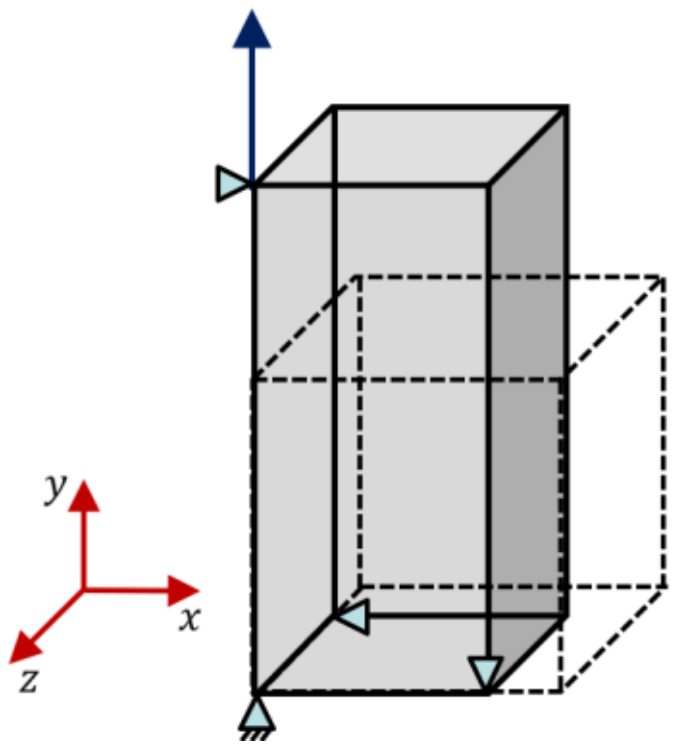

```
{
    {"vertex_list": ["V000"],
     "constraints": ["free", "free", "fixed"],
     "loading_type": None,
     "loading_mode": None,
     "applied_load": {}
    },
    {"vertex_list": ["V001"],
     "constraints": ["fixed", "fixed", "fixed"],
     "loading_type": None,
     "loading_mode": None,
     "applied_load": {}
    },
    {"vertex_list": ["V101"],
     "constraints": ["free", "fixed", "free"],
     "loading_type": None,
     "loading_mode": None,
     "applied_load": {},
    },
    {"vertex_list": ["V011"],
     "constraints": ["fixed", "loaded", "free"],
     "loading_type": "force",
     "loading_mode": "static",
     "applied_load": [{"magnitude": 100,
                       "frequency": 0,
                       "duration": 250,
                       "R": 0}]
    },
    {"vertex_list": ["V111"],
     "constraints": ["free", "free", "free"],
     "loading_type": None,
     "loading_mode": None,
     "applied_load": {}
     }
}
```

(b)

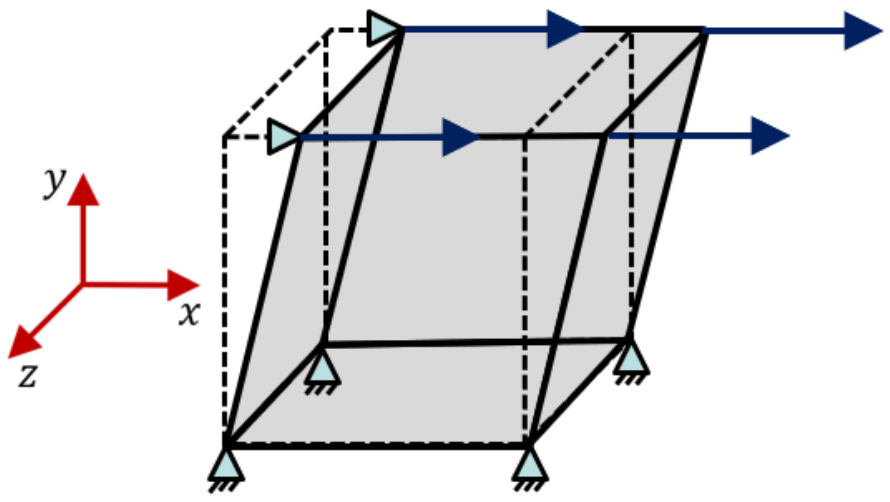

```
{
    {"vertex_list": ["V000", "V001", "V101", "V100"],
     "constraints": ["fixed", "fixed ", "fixed"],
     "loading_type": None,
     "loading_mode": None,
     "applied_load": {}
    },
    {"vertex_list": ["V010", "V011", "V111", "V110"],
     "constraints": ["loaded", "free", "free"],
     "loading_type": "force",
     "loading_mode": "static",
     "applied_load": [{"magnitude": 100,
                       "frequency": 0,
                       "duration": 250,
                       "R": 0}]
    },
    {"vertex_list": ["V011", "V010"],
     "constraints": ["fixed", "free", "free"],
     "loading_type": None,
     "loading_mode": None,
     "applied_load": {},
    }
}
```

**Figure 4**. (a) An example of periodic mechanical boundary conditions applied on representative vertices of an RVE showing a simple tensile loading in the y-direction; in this case, the vertex list includes only a single vertex. (b) An example of mechanical boundary conditions applied to an RVE under simple shear loading in the x-direction. Two faces normal to the y-direction are fixed and subjected to loading, and one edge parallel to the z-direction is fixed. A list containing four vertices represents a face, while a list containing two vertices represents an edge.

### 4.3.3. Material model

Within this metadata schema, we aim to create a generic and specific framework and design elements that effectively capture information, ensuring we have a FAIR data object. This facilitates sharing and collaboration, even when workflows are external to the system. Although the elements are designed to be generic, the modular nature of this schema allows for complete customization based on specific needs. This flexibility is particularly important for *the material model* category due to the many different possibilities and complexities involved.

In realistic micromechanical modeling, each material behaves differently due to its unique characteristics defined by different constitutive models. There are realistic scenarios where, within a Representative Volume Element (RVE), more than one phase exists, and this description must exist for each phase explicitly. Thus, this material model would include all available phases with their own characteristics. In this schema, the material can be described within three categories: a name or identifier of the phase, the constitutive model, and the crystallographic orientations of the grains. The phase identifier serves as a unique string that identifies each phase, ensuring that every entry is distinct and easily referable. The constitutive model explicitly details material behavior in three main categories: elastic, plastic and damage, as shown in Table 6. Within each category, we do not aim to impose fixed element names, as doing so would reduce the flexibility and usefulness of the schema. Instead, we offer a general guideline, allowing users to describe any constitutive model. Furthermore, we emphasize that specifying common names for the constitutive models and parameters will increase the usefulness of a digital object as it allows for specific searches and thus significantly improves the findability of the digital object. The elastic parameters can include material-specific parameters like Young's modulus or Poisson's ratio in the simplest case and C11, C12, and C44 as stiffness matrix components in more complex anisotropic scenarios. In the plastic category, all the material-specific parameters related to how the material responds plastically, including the model's name and the material-specific parameters, are captured. For instance, the parameters of a J2 plasticity model might include the initial yield strength and anisotropic material coefficient. They may involve additional hardening models like Voce, which can be described with two material-specific parameters. In more complex scenarios like the crystal plasticity model, the plastic parameters include slip planes and critical resolved shear stress (CRSS). Moreover, if extra hardening models such as Armstrong-Frederick [34], Chaboche [35], or Ohno-Wang [36] are involved, their material-specific parameters should be included in the plastic parameters category. The damage model category in the schema allows the addition of various damage models and their corresponding parameters, enabling the representation of material degradation and failure behaviors. This structure ensures flexibility and comprehensiveness, allowing for the inclusion of various constitutive models and their associative parameters. **Figure 5** shows an example of the constitutive model section, including J2 plasticity combined with Voce hardening.

```
{
    "elastic_model_name": "Linear Elasticty"
    "elastic_parameters": {"youngs_modulus": 210000,
                           "poissons_ratio": 0.3}
    "plastic_model_name": "J2 Plasticity with Voce Hardening"
    "plastic_parameters": {"initial_yield_stress": 250,
                           "voce_saturation_stress": 350,
                           "voce_hardening_rate": 0.05}
    "units": {"Stress": "MPa", "Stiffness": "MPa"}
}
```

**Figure 5**. Constitutive model element for J2 Plasticity with Voce Hardening. Refer to **Appendix 1** for more details.

Also, the crystallographic orientation of all grains, i.e., the material's texture, must be included and captured within the orientation category. This field can remain empty in scenarios where the material behavior is isotropic, as orientation information is not necessary for isotropic materials. The crystallographic texture can be captured with three sets of Euler angles assigned to each grain of the material. The specific definition of the frame, i.e., lab or material frame, and active or passive rotation should also be provided when possible. In many cases, searching for specific textures is essential. We recommend assigning a unique identifier for each orientation to avoid searching within arrays of orientations. This can be achieved by using a hashing technique to translate arrays of Euler angles into unique identifiers. The elements in this category can be seen in **Table 6**.

**Table 6**. Material-specific elements, obligation, type and property, and description

<table>
<tr><th>Components</th><th>Obl</th><th>Type</th><th>Property</th><th>Description</th></tr>
<tr><td>material_identifier</td><td>M</td><td>Str</td><td>-</td><td>Name identifying the material</td></tr>
<tr><td>constitutive_model</td><td>M</td><td>Obj</td><td>-</td><td>
<table>
<tr><th>Components</th><th>Obl</th><th>Type</th><th>Property</th><th>Description</th></tr>
<tr><td>elastic_model_name</td><td>O</td><td>Obj</td><td>-</td><td>Name of the elastic model.</td></tr>
<tr><td>elastic_parameters</td><td>O</td><td>Obj</td><td>-</td><td>Elastic properties of the material.</td></tr>
<tr><td>plastic_model_name</td><td>O</td><td>Str</td><td>-</td><td>Name of the plasticity model (including hardening).</td></tr>
<tr><td>plastic_parameters</td><td>O</td><td>Obj</td><td>-</td><td>Parameters specific to the plasticity model.</td></tr>
<tr><td>damage_model_name</td><td>O</td><td>Str</td><td>-</td><td>Name of the damage model.</td></tr>
<tr><td>damage_parameters</td><td>O</td><td>Obj</td><td>-</td><td>Parameters specific to the damage model.</td></tr>
</table>
</td></tr>
<tr><td>orientation</td><td>O</td><td>Obj</td><td>-</td><td>
<table>
<tr><th>Components</th><th>Obl</th><th>Type</th><th>Property</th><th>Description</th></tr>
<tr><td>euler_angles</td><td>M</td><td>Obj</td><td>-</td><td>The orientation of each grain given as Euler angles in active rotation from global coordinates.
<table>
<tr><th>Components</th><th>Obl</th><th>Type</th><th>Property</th></tr>
<tr><td>Phi1</td><td>M</td><td>Array of Floats</td><td>Angle</td></tr>
<tr><td>Phi</td><td>M</td><td>Array of Floats</td><td>Angle</td></tr>
<tr><td>Phi2</td><td>M</td><td>Array of Floats</td><td>Angle</td></tr>
</table>
</td></tr>
<tr><td>grain_count</td><td>M</td><td>Int</td><td>-</td><td>The total number of material Grains equal to the number of orientation angles.</td></tr>
<tr><td>orientation_identifier</td><td>O</td><td>Str</td><td>-</td><td>Hash or identifier for the orientation.</td></tr>
<tr><td>texture_type</td><td>M</td><td>Str</td><td>-</td><td>Free-text description of texture type.</td></tr>
<tr><td>frame</td><td>O</td><td>Str</td><td>-</td><td>The coordinate system used for the orientation data. Can be either "lab" (fixed relative to laboratory) or "material" (fixed relative to material).</td></tr>
<tr><td>rotation_type</td><td>O</td><td>Str</td><td>-</td><td>Specifies whether the rotation is described as "active" (physical rotation of the grain) or "passive" (rotation of the coordinate system)</td></tr>
<tr><td>software</td><td>O</td><td>Str</td><td>-</td><td>Name of software or tools used to generate orientation</td></tr>
<tr><td>software_version</td><td>O</td><td>Str</td><td>-</td><td>The version of the software used to generate orientation</td></tr>
</table>
</td></tr>
</table>

### 4.4. Property elements

A post-processing step is usually used to extract the macroscopic behavior from the simulation results within a micromechanical workflow. This step usually involves translating detailed microstructural behavior into macroscopic material properties through homogenization techniques of the field variables. Through homogenization techniques, a small statistically representative sample of the material is used to derive effective continuum properties. By averaging the behavior over the microscale, these techniques allow for the prediction of the material's overall behavior at the macroscale. The detailed structure of the property elements can be seen in **Table 7**. It must be noted that the property components can be extended when needed. Additionally, the stress and strains are described in the table using their six independent components of the tensor, which is the general case. However, this is not limiting and can be modified based on availability, requirements, or the need to have only equivalent values.

**Table 7.** Property elements, obligation, type, property, and description.

| Components | Obl | Type | Description: Components | Type | Obl | Property | Description |
|---|---|---|---|---|---|---|---|
| stress | M | Obj | equivalent_stress | Array of Floats | M | Stress | List of equivalent stress values throughout the simulation. |
| | | | stress_11 | Array of Floats | O | Stress | list of stress values in direction 11 throughout the simulation in the global coordinate system |
| | | | stress_22 | Array of Floats | O | Stress | list of stress values in direction 22 throughout the simulation in the global coordinate system |
| | | | stress_33 | Array of Floats | O | Stress | list of stress values in direction 33 throughout the simulation in the global coordinate system |
| | | | stress_12 | Array of Floats | O | Stress | list of stress values in direction 12 throughout the simulation in the global coordinate system |
| | | | stress_13 | Array of Floats | O | Stress | list of stress values in direction 13 throughout the simulation in the global coordinate system |
| | | | stress_23 | Array of Floats | O | Stress | list of stress values in direction 23 throughout the simulation in the global coordinate system |
| total_strain | M | Obj | equivalent_strain | Array of Floats | M | Strain | List of equivalent strain values throughout the simulation |
| | | | strain_11 | Array of Floats | O | Strain | list of total strain values in direction 11 throughout the simulation in the global coordinate system |
| | | | strain_22 | Array of Floats | O | Strain | list of total strain values in direction 22 throughout the simulation in the global coordinate system |
| | | | strain_33 | Array of Floats | O | Strain | list of total strain values in direction 33 throughout the simulation in the global coordinate system |
| | | | strain_12 | Array of Floats | O | Strain | list of total strain values in direction 12 throughout the simulation in the global coordinate system |
| | | | strain_13 | Array of Floats | O | Strain | list of total strain values in direction 13 throughout the simulation in the global coordinate system |
| | | | strain_23 | Array of Floats | O | Strain | list of total strain values in direction 23 throughout the simulation in the global coordinate system |
| plastic_strain | O | Obj | equivalent_plastic_strain | Array of Floats | O | Strain | List of equivalent plastic strain values throughout the simulation |
| | | | plastic_strain_11 | Array of Floats | O | Strain | list of plastic strain values in direction 11 throughout the simulation in the global coordinate system |
| | | | plastic_strain_22 | Array of Floats | O | Strain | list of plastic strain values in direction 22 throughout the simulation in the global coordinate system |
| | | | plastic_strain_33 | Array of Floats | O | Strain | list of plastic strain values in direction 33 throughout the simulation in the global coordinate system |
| | | | plastic_strain_12 | Array of Floats | O | Strain | list of plastic strain values in direction 12 throughout the simulation in the global coordinate system |
| | | | plastic_strain_13 | Array of Floats | O | Strain | list of plastic strain values in direction 13 throughout the simulation in the global coordinate system |
| | | | plastic_strain_23 | Array of Floats | O | Strain | list of plastic strain values in direction 23 throughout the simulation in the global coordinate system |

### 4.5. Units

Besides the main categories described above, the schema includes a units section that standardizes the property units. All the metadata elements mentioned previously are linked to a property, where applicable. If a property has been defined for an element, the corresponding units can be found in the units section. This ensures proper data usage and can increase interpretation of the data. The detailed structure of the unit element can be seen in **Table 8**.

**Table 8**. Unit elements, type, and value

| Components | Type | Value | Conversion factor to SI unit |
|---|---|---|---|
| Stress | Str | MPa | $10^6$ |
| Strain | Int | 1 | 1 |
| Stiffness | Str | MPa | $10^6$ |
| Length | Str | mm | $10^{-3}$ |
| Angle | Str | Radian | 1 |
| Temperature | Str | Kelvin | 1 |
| Force | Str | N | 1 |

## 5. A Use Case of the Metadata Schema

This section demonstrates the practical application of the designed metadata schema through a specific use case of a crystal plasticity finite element method simulation. The material under investigation is copper in form of a single-phase polycrystalline structure. The geometry is defined using a 3D cubic RVE comprising 343 grains which can be seen in **Figure 6**. Each grain is characterized by a distinct orientation, defined by three Euler angles, depicting a severely anisotropic Goss texture. A crystal plasticity constitutive model was employed to capture the anisotropic mechanical behavior. The elastic behavior was defined using three stiffness constants, and the plastic behavior was characterized using five material-specific parameters critical for modeling the slip mechanisms and hardening effects inherent in crystal plasticity. The orientation was generated using the MTEX tool in MATLAB. To enhance calculation precision, each grain is discretized into eight finite elements. The periodic boundary conditions are applied to four representative nodes as monotonic increasing stress to simulate a realistic material behavior. The Abaqus software was used to solve the boundary value problems. The homogenized stress accumulated plastic strain and total strain were calculated through a post-processing step. More details can be found in [37].

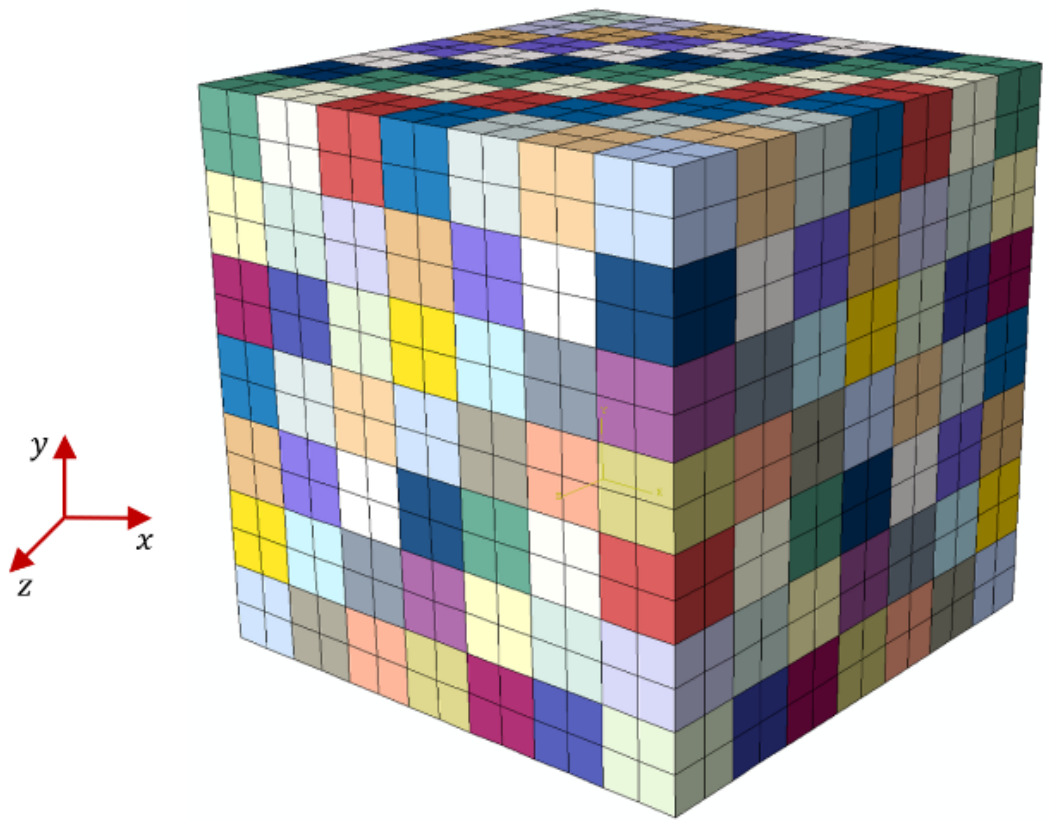

**Figure 6.** A three-dimensional RVE used for the CPFEM simulation of a polycrystalline material. The RVE has dimensions of 0.665 x 0.665 x 0.665 millimeters and is composed of 343 grains. Each grain is assigned a distinct crystallographic orientation, contributing to the severely anisotropic Goss texture. It is important to note that the colors representing the grains are chosen arbitrarily and do not correspond to specific grain orientations. Each grain is subdivided into eight finite elements, resulting in a total of 2744 elements. Each element has a size of 0.095 x 0.095 x 0.095 millimeters, and the discretization type is structured.

The generated data object has been structured through the designed metadata schema and includes all the necessary information to describe the data as a FAIR data object. The data object is available online [38]. As explained in section 4.1, providing a unique identifier or name for the data object is very important. One method we have used is creating a hash from all available mandatory data object elements. A hash function takes an input and returns a fixed-size string of bytes. In the template script, the HD5 hash function is used for this task. The output is typically a unique sequence of letters and numbers. Even a small change in the input will produce a significantly different hash, making this method a good possibility for generating unique identifiers. This technique is very fast and deterministic, meaning the same input will always produce the same output hash.

## 6. Discussion and outlook: pathway to a sustainable data ecosystem

Even though we have defined the data object using the designed metadata schema to meet all the requirements for being a FAIR information entity, realizing its full potential requires further steps. The metadata schema is designed to ensure all the information necessary for defining a FAIR data object is collected and handled properly. However, without a proper database platform, this potential remains theoretical. A dedicated database platform is essential to managing, storing, and retrieving data objects effectively. Therefore, while we have successfully designed a metadata schema to ensure having FAIR data objects of mechanical materials data, handling many of these data objects within a sustainable data ecosystem requires a proper database infrastructure. This step is crucial for supporting the full life cycle of these data objects, ensuring their value, sustainability, and usability for future research.

The workflow-centric approach presented in this work can play a pivotal role in building a sustainable data ecosystem. By systematically organizing metadata according to workflow steps, this method not only ensures FAIR compliance but also simplifies the process of expanding and adapting the schema to future workflows. This flexibility is essential for sustainability as research practices and technologies evolve over time. The scalability and adaptability of this approach to different workflows, whether computational or experimental, make it a strong foundation for building a sustainable, long-term data ecosystem. Furthermore, the automation of workflows provides an additional layer of sustainability. As workflows are automated, the process of extracting, retrieving, and integrating metadata elements becomes more efficient, reducing the burden on researchers. This automation minimizes manual errors and makes it easier to maintain the quality and consistency of metadata across different datasets, ensuring that data objects remain findable, accessible, interoperable, and reusable (FAIR). The adoption of such an automated, workflow-centric system is an essential step toward creating a sustainable, efficient, and user-friendly data ecosystem.

Assessing the FAIRness of individual data objects or the data ecosystem as a whole can be done through different methods. Fair principles themselves provide a high-level guideline, but the specific criteria and metrics for assessing date FAIRness should be developed by domain experts. Engaging the scientific community through workshops and user surveys is essential for developing and refining the criteria for assessing FAIRness. These methods gather insights and feedback from the individuals who will use and interact with the data ecosystem, ensuring that the system meets their needs and expectations. Different community members might interact with the data ecosystem in various ways and have different needs. Some examples are given in **Table 9**. A good FAIR assessment involves considering the specific needs of different users and ensuring that the data object is FAIR enough to satisfy all those needs. This approach is hypothetical and can include completely different scenarios and members, but this diversity in member feedback provides valuable insights and can lead to improvement in existing metadata schema.

**Table 9**. Feedback and Assessment Scenarios

| Community Member | Specific Needs |
|---|---|
| Computational Mechanics Expert | A material scientist's expert wants to analyze the mechanical properties of a specific material. The person should understand the material definition, the model used, and applied boundary conditions. |
| Material Scientist Expert | An expert from a closely related field like materials science wants to compare their experimental data with this data to find the correlation. They should be able to get the information from the metadata elements. |
| Software Developer | A developer who wants to create a web application that visualizes mechanical data in correlation with microstructural information needs access to both mechanical properties data and all the relevant metadata. They must be self-explanatory as the person is not a field expert. |
| Engineer | An engineer wants to reproduce the results of micromechanical simulation with another software like ANSYS. They should be able to rely only on the metadata and produce comparable results. |

## 7. Conclusion

This work aimed to address data sparsity in the field of materials mechanics by proposing a workflow-centric approach to generating FAIR data objects for mechanical properties. Recognizing the strong microstructure-sensitivity and history dependence of mechanical properties, our objective was to develop a metadata schema that integrates both metadata and mechanical data, ensuring that the resulting data objects adhere to FAIR principles. This effort contributes to higher-level goals of sustainable and effective research data management, addressing a critical need in the computational mechanics field with its vast amounts of research data.

The data object can be defined as a unique instance of any simulation workflow. The data object can be described as a FAIR data object by designing elements that belong to four different categories, including user, system, job and property information. User-level information includes general elements that can be extracted automatically from the environment as well as elements requiring active user decisions. Job-related elements, the most specific level, are further categorized into geometry, material model, and boundary conditions using a functional taxonomy. When designing these elements, a strong emphasis was put on their simplicity, extensibility, and flexibility. This approach ensures that the schema is not only easy to implement and to maintain but it is also capable of adapting to future needs, thus promoting long-term sustainability and effective data management. For user and system-level elements, we adopted components from generally accepted schemas and performed crosswalks. For job-specific elements, where no equivalent schema existed, we defined descriptive and understandable element names to ensure community interoperability. As a showcase, we used our metadata schema to describe a FAIR data object from a micromechanical simulation workflow. In this example, the material was copper, represented by a cubic RVE, and a crystal plasticity finite element method was used to study the stress-strain response under a complex multiaxial loading scenario. It is important to note that while this work represents a significant step in defining and implementing a metadata schema for microstructure-sensitive mechanical properties, it currently focuses exclusively on computational workflows, particularly micromechanical simulations. However, the workflow-centric approach can serve as a general guideline for structuring metadata across any type of workflow. Our approach gathers the necessary information based on the specific steps within the workflow. We believe that this approach can significantly aid in describing data as a FAIR data object, ensuring that all relevant elements are captured as part of the workflow itself. While we cannot provide a specific schema for every possible workflow, the underlying concept allows the approach to be applied broadly, provided the workflow is well-defined. The strength of this approach lies in its ability to guide the user in identifying where to find the necessary information to describe the data as FAIR as possible. As workflows are often automated, the process of extracting and retrieving metadata becomes straightforward, enabling users to integrate the required elements efficiently. This flexibility allows the schema to be applied to both computational and experimental workflows.

It should be noted that while a major step has been taken in defining and implementing a metadata schema for microstructure-sensitive mechanical properties, the journey toward a sustainable data ecosystem is ongoing. Establishing a robust database platform for mechanical materials data is essential for realizing the full potential of these efforts, ensuring efficient management, storage, and retrieval of data objects. Additionally, continuous engagement within the scientific community to assess and refine data objects through FAIR assessments can ensure that the data ecosystem remains relevant.

**Conflict of Interest**

The authors declare no conflict of interest.

**Supporting Information**

The data that support the findings of this study is openly available on Zenodo at https://doi.org/10.5281/zenodo.13885748 [38]. The JSON schema, which includes all mandatory and optional elements with detailed explanations, along with a Python script to generate data objects and create a unique identifier for each, are available in the GitHub repository: https://github.com/Ronakshoghi/MetadataSchema

**Acknowledgments**

AH gratefully acknowledges funding by the NFDI consortium NFDI-MatWerk in the context of the work of the association German National Research Data Infrastructure (NFDI) e.V. NFDI is financed by the Federal Republic of Germany and the 16 federal states and funded by the Federal Ministry of Education and Research (BMBF) – funding code M532701 / the Deutsche Forschungsgemeinschaft (DFG, German Research Foundation) under the National Research Data Infrastructure – NFDI 38/1 – project number 460247524.

Received:

Revised:

Published online

**Appendix 1**

As explained in section 4.3.3, the constitutive_model element is defined as an object that includes several categories—such as elastic, plastic, and damage behaviors. These categories are designed to allow the inclusion of appropriate parameters depending on the specific use case. The model's structure is not fixed, providing flexibility to define custom constitutive models tailored to different material behaviors. A general template, available online at https://github.com/Ronakshoghi/MetadataSchema/blob/main/general_constitutive_model_metadata_schema.json , serves as a foundation for this approach, allowing users to inherit the core structure while defining any specific custom constitutive model, which can then be included in the data object. This approach ensures that essential components like elastic, plastic, and damage behaviors are organized in a standardized way while still allowing flexibility to define additional properties as needed. The schema's modularity allows users to inherit the general structure and extend it to build specific constitutive models.

For instance, in the case of the J2 plasticity model with Voce hardening, which is also available online at https://github.com/Ronakshoghi/MetadataSchema/blob/main/J2_plasticity_voce_hardening_constitutive_model_metadata_schema.json , we designed our custom schema by adding specific elastic and plastic properties without altering the main structure. We filled in the elastic_parameters with "youngs_modulus" and "poissons_ratio" to describe the material's elastic behavior and added "initial_yield_stress", "voce_saturation_stress", and "voce_hardening_rate" to the plastic_parameters. By doing so, we customized the model based on our specific needs while maintaining the overall structure provided by the general schema.

To assist users in creating their own constitutive models, Table X provides a set of suggested element names that can be used to represent key material parameters. This table offers a helpful guide for defining material parameters in a standardized way while allowing for flexibility when developing new constitutive models based on the modular general schema.

| Component | Symbol | Origin | Element Name | Property |
|---|---|---|---|---|
| Youngs Modulus | $E$ | Linear Elasticity | youngs_modulus | Stiffness |
| Shear Modulus | $G, \mu$ | Linear Elasticity | shear_modulus | Stiffness |
| Elastic Constant | $C11$ | Anisotropic Elasticity | C11 | Stiffness |
| Elastic Constant | $C12$ | Anisotropic Elasticity | C12 | Stiffness |
| Elastic Constant | $C44$ | Anisotropic Elasticity | C44 | Stiffness |
| Poisson's ratio | $\nu$ | Linear Elasticity | poissons_ratio | - |
| Initial Yield Stress | $\sigma_{y0}$ | J2 Plasticity | initial_yield_stress | Stress |
| Voce Saturation Stress | $\sigma_s$ | Voce Hardening [39] | voce_saturation_stress | Stress |
| Voce Hardening Parameter | $b$ | Voce Hardening | voce_hardening_rate | - |
| Hollomon Strength Coefficient | $C$ | Hollomon Model [40] | hollomon_strength_coefficient | Stress |
| Hollomon Hardening Rate | $n$ | Hollomon Model | hollomon_hardening_rate | - |
| Ludwik Strength Coefficient | $C$ | Ludwik Hardening [41] | ludwik _strength_coefficient | Stress |
| Ludwik Hardening Rate | $n$ | Ludwik hardening | ludwik_hardening_rate | - |
| Swift Strength Coefficient | $C$ | Swift hardening [42] | swift_strength_coefficient | Stress |
| Swift Hardening Rate | $n$ | Swift hardening | swift_hardening_rate | - |
| Johnson-Cook Hardening Modulus | $B$ | Johnson-Cook Model [43] | johnsoncook_hardening_modulus | Stress |
| Johnson-Cook Hardening Rate | $n$ | Johnson-Cook Model | johnsoncook_hardeningrate | - |
| Johnson-Cook Strain Rate Sensitivity | $C$ | Johnson-Cook Model | johnsoncook_strain_rate_sensetivity | - |
| Johnson-Cook Reference Temperature | $T_0$ | Johnson-Cook Model | johnsoncook_reference_tempearture | Temperature |
| Johnson Cook Temperature exponent | $m$ | Johnson-Cook Model | johnson_cook_temperature_exponent | - |
| Melting Temperature | $T_m$ | Johnson-Cook Model | melting_temperature | Temperature |
| Chaboche Hardening Modulus | $C_i$ | Chaboche Model [44] | chaboche_hardening_modulus_i | Stress |
| Chaboche Dynamic Recovery Rate | $\gamma_i$ | Chaboche Model | chaboche_dynamic_recovery_rate_i | - |
| Ohno-Wang Hardening Modulus | $C_i$ | Ohno-Wang Model | ohnowang _hardening_modulus_i | Stress |
| Ohno-Wang Dynamic Recovery Rate | $\gamma_i$ | Ohno-Wang Model | ohnowang_dynamic_recovery_rate_i | - |
| Ohno-Wang Effective Modulus | $\eta_i$ | Ohno-Wang Model | ohnowang_effective_modulus_i | Stress |
| Armstrong-Frederick Hardening Modulus | $C$ | Armstrong-Frederick Model [34] | armstrongfrederick _hardening_modulus | Stress |
| Armstrong-Frederick Dynamic Recovery Rate | $\gamma$ | Armstrong-Frederick Model | armstrongfrederick_hardening_modulus | - |
| Ramberg-Osgood Strength Coefficient | $K$ | Ramberg-Osgood Model [45] | rambergosgood_strength_coefficient | Stress |
| Ramberg-Osgood Hardening Rate | $n$ | Ramberg-Osgood Model | rambergosgood_hardening_rate | - |
| Ramberg-Osgood Constant | $\alpha$ | Ramberg-Osgood Model | rambergosgood_constant | - |
| Reference Shear Rate | $\dot{\gamma}_0$ | Crystal Plasticity [46] | reference_shear_rate | - |
| Initial Critical Resolved Shear Stress | $\tau_0$ | Crystal Plasticity | initial_critical_resolved_shear_stress | Stress |
| Saturated slip resistance | $\tau_s$ | Crystal Plasticity | saturated_slip_resistance | Stress |
| Strain Rate Sensitivity Exponent | $m$ | Crystal Plasticity | strain_rate_sensitivity_exponent | - |
| Reference Hardening Rate | $h_0$ | Crystal Plasticity | reference_hardening_rate | Stress |
| Hardening Exponent | $p$ | Crystal Plasticity | hardening_exponent | - |
| Interaction Strength | $\chi^{\alpha\beta}$ | Dislocation-Based Crystal plasticity [47] | interation_strength | - |
| Density of Mobile Dislocation | $\varrho_m^\alpha$ | Dislocation-Based Crystal plasticity | mobile_dislocation_desnity | Density |
| Burgers Vector Magnitude | $b$ | - | burgers_vector_magnitude | Length |
| Mobile Dislocation Velocity | $\nu^\alpha$ | Dislocation-Based Crystal plasticity | mobile_dislocation_velocity | Velocity |
| Dislocation Density Constant | $c_i$ | Dislocation-Based Crystal plasticity | dislocation_desnity_constant_i | - |
| Dislocation Jump width | $\lambda^\alpha$ | Dislocation-Based Crystal plasticity | dislocation_jump_width | Length |
| Dislocation Attack Frequency | $\nu_{attack}$ | Dislocation-Based Crystal plasticity | dislocation_attack_frequency | Frequency |
| Effective Glide Activation Energy | $Q_{slip}$ | Dislocation-Based Crystal plasticity | effective_glide_activation_energy | Energy |
| Glide Activation Volume | $V_{glide}$ | Dislocation-Based Crystal plasticity | glide_activation_volume | Volume |
| Critical Dipole Formation Distance | $d_{dipole}$ | Dislocation-Based Crystal plasticity | critical_dipole_formation_distance | Length |
| Diffusion Coefficient | $D_0$ | Dislocation-Based Crystal plasticity | diffusion_coefficient | Diffusion Coefficient |
| Geometrical Factor | $c_1$ | Geometrically Necessary Dislocation (GND) Density Model [48] | geometrical_factor | - |
| Average Dislocation Pile-Up Size | $L$ | Geometrically Necessary Dislocation (GND) Density Model | average_dislocation_pileup_size | Length |

## Appendix 2

The "*shared_with*" element which was described in section 4.1 (User-specific elements) specifies who can access the data object. Access control is crucial for ensuring data security, proper sharing, and collaboration. This appendix provides detailed explanations and examples for each category used for this element.

**Categories:**

- c (Creator)

The creator of the data object. The creator is responsible for the generation and initial ownership of the data. There is no need to specify the name/s as the creator element already exists.

- Access Rights: Full access to the data.
- Example Usage: "c"
- Scenario: Only the creator can view, edit, and retrieve the data.

- u (User)

Specific users who are granted access to the data object. Each user must be specified by name. The name is formatted as “Last name, First name”. Multiple users are given in a list.

- Access Rights: Defined by the creator or data manager; typically includes viewing, editing, or specific data manipulation permissions.
- Example Usage: ["Doe, John", "Smith, Jane"]
- Scenario: The users John Doe and Jane Smith are granted access to the data.

- g (Group)

A specific group of individuals within an organization or project team who are granted access to the data object. The group's name must be specified.

- Access Rights: Typically includes permissions for all group members to view and edit.
- Example Usage: "Micromechanical and Macroscopic Modelling"
- Scenario: All " Micromechanical and Macroscopic Modelling" members can access the data, facilitating group collaboration and shared responsibilities.

- all (Public Access)

Public access. Making the data available to everyone without restriction.

- Access Rights: Full public visibility; anyone can view (and potentially edit, depending on the dataset's configuration) the data.
- Example Usage: "all"
- Scenario: The data is publicly available for broader community use, open research, and transparency.

- Mixed categories

A combination of different access categories to specify multiple entities that can access the data object.

- Example: "shared_with": ["c", ["Doe, John", "Smith, Jane"], ["Micromechanical and Macroscopic Modelling"]]
- Scenario: besides the creator, the data object is shared with two users and one group.